# MCP GUARDIAN: A SECURITY-FIRST LAYER FOR SAFEGUARDING MCP-BASED AI SYSTEM

Sonu Kumar [1], Anubhav Girdhar [2], Ritesh Patil [3], Divyansh Tripathi [4], Nishanth Veduruvada [5], Madhukar Anugu [6], Sneha Roy [7], Venkata Talatam [8]

[1] R&D, Sporo Health, USA.
[2] Data Engineering and AI, Involead, India.
[3] Gen AI CoE, Capgemini, India.
[4] IIT Roorkee, India.
[5] Arc Steam Technologies, India
[6] Software Engineering, H&R Block, USA
[7] CoRe DDAI, Deloitte USI, India
[8] AI Engineering, Kodryx AI, India

***ABSTRACT***

*As Agentic AI gain mainstream adoption, the industry invests heavily in model capabilities, achieving rapid leaps in reasoning and quality. However, these systems remain largely confined to data silos, and each new integration requires custom logic that is difficult to scale. The Model Context Protocol (MCP) addresses this challenge by defining a universal, open standard for securely connecting AI-based applications (MCP clients) to data sources (MCP servers). However, the flexibility of the MCP introduces new risks, including malicious tool servers and compromised data integrity. We present MCP Guardian, a framework that strengthens MCP-based communication with authentication, rate-limiting, logging, tracing, and Web Application Firewall (WAF) scanning. Through real-world scenarios and empirical testing, we demonstrate how MCP Guardian effectively mitigates attacks and ensures robust oversight with minimal overheads. Our approach fosters secure, scalable data access for AI assistants, underscoring the importance of a defense-in-depth approach that enables safer and more transparent innovation in AI-driven environments.*

## 1. INTRODUCTION

LLMs have witnessed a rapid expansion in both scale and capability, demonstrating unprecedented performance in tasks ranging from natural language generation to complex programming challenges. While initially confined to relatively passive roles—delivering text-based answers or summaries—LLMs are now increasingly being placed in “agentic” positions, where they not only generate content but also initiate and orchestrate actions across various external systems. This paradigm shift underscores how LLMs can serve as decision-making engines, interfacing with diverse tools, such as databases, web services, and file systems. By autonomously chaining multiple tool calls, agentic workflows can solve sophisticated problems that extend beyond the written word.

However, unlocking these extended capabilities has introduced significant engineering complexity, largely because of the lack of standardized interfaces. Historically, developers resorted to custom “plugin” or “adapter” logic for each new external tool, leading to fragmented solutions that are difficult to maintain at scale. To address this fragmentation, the Model Context Protocol (MCP) was recently proposed as a universal “multiplexer,” enabling LLM-powered clients to discover and invoke tool servers in a unified manner. By abstracting the underlying implementation details, the MCP simplifies tool integration, thereby lowering the barrier to building AI applications that can incorporate external data and services.

However, this newfound flexibility comes with an increased risk. LLMs or more advanced agentic workflows that can autonomously access file systems or databases pose non-trivial security challenges: a maliciously crafted prompt or compromised server can result in unauthorized data exfiltration, destructive operations, or other exploitative behaviors. Moreover, this agentic paradigm requires enhanced observability. Traditional logging and monitoring methods are insufficient for capturing the complex chains of reasoning and actions that an LLM may perform when orchestrating multiple tools in parallel. The absence of thorough instrumentation complicates both real-time auditing and post-hoc forensics, raising concerns about transparency and compliance.

In light of these challenges, this study introduces MCP Guardian, a comprehensive middleware layer aimed at securing and monitoring the interactions between MCP Clients and MCP-based tool servers. Drawing inspiration from zero-trust security frameworks, web application firewalls, and distributed tracing practices, MCP Guardian intercepts every tool call to:

- Enforce authentication and authorization checks,
- Apply rate-limiting strategies to protect against abuse or runaway processes,
- Provide extensive logging and tracing for transparent auditing, and
- Scan suspicious input patterns via a lightweight Web Application Firewall (WAF).

This contribution synthesizes insights from LLM alignment, software security, and distributed system observability to propose a practical solution for the next generation of AI-driven agents. In particular, we highlight the following points:

1. **Problem Analysis:** A thorough examination of security and observability gaps in MCP-based systems, especially where LLMs autonomously issue tool calls.
2. **Framework Design:** A detailed architectural description of MCP Guardian highlighting its core components (authentication, access control, request logging, rate limiting, and WAF scanning) and how they interoperate.
3. **Implementation and Evaluation:** A reference implementation in Python, tested on real-world scenarios, including a weather-tool MCP server, to illustrate both security efficacy and performance overhead.
4. **Empirical and Theoretical Insights:** Scenario-based testing of malicious inputs, latency measurements, and throughput analyses, offering an understanding of how MCP Guardian scales and adapts to various domains.

# 2. LITERATURE REVIEW

## 2.1. AI Agents and Tool Integration

Recent scholarly interest in AI agents has intensified, driven by the desire to move beyond passive text generation and empower Large Language Models (LLMs) to autonomously perform tasks in real-world contexts. Early attempts at "tool use" often relied on bespoke plugins or direct calls to specialized APIs. For instance, OpenAI's ChatGPT introduced plugin frameworks that connect to external services[3], while other AI-based "copilot" tools were designed to read and write files in code repositories. Despite these innovations, the lack of a unified, standardized method for discovering and invoking tools frequently forced developers to create patchwork solutions, thereby increasing the risk of security vulnerabilities, inconsistent access controls, and a limited audit trail.

## 2.2. The Emergence of MCP

The Model Context Protocol (MCP)—promoted by Anthropic[1] and further explored by others [2]—addresses these integration challenges by offering an open, extensible protocol for LLM-driven interactions with external tools. By allowing AI clients to query a server for available functions and associated metadata, MCP significantly reduces the repeated overhead encountered in ad-hoc "plugin" models. Instead of requiring specialized integrations for each tool, a single request/response channel (e.g., JSON over stdio or HTTP) serves as a universal interface. This design shares similarities with gRPC or JSON-RPC but is optimized for LLMs' iterative reasoning, where multiple tool calls may be chained in a single session.

However, MCP's openness also presents a notable attack surface. Malicious or compromised MCP servers can cloak harmful code under seemingly benign functionality and exfiltrate sensitive data. Consequently, security researchers and practitioners have highlighted the need for robust authentication, authorization, and supply chain protections within the MCP ecosystem. Some preliminary guidelines exist for "trusted" AI-tool environments[4], but comprehensive security strategies remain underdeveloped for broader, community-driven MCP deployments.

## 2.3. Security in LLM-Orchestrated Workflows

As LLMs evolve into autonomous agents, the scope of potential security risks widens substantially. These systems can, in principle, read, write, and execute code, posing serious threats if not rigorously controlled. Zero-trust architectures emphasize continuous request validation rather than assuming any "trusted" status at the outset, an approach that applies naturally to MCP: every incoming request to a tool server should undergo authentication, parameter sanitization, and logging.

Li and Hsu[2] underscore the importance of policy-driven authorization in agentic AI, where a model's capacity to decide which actions to take must align with stringent security and compliance policies. In tandem, DevSecOps best practices mandate code scanning, secret rotation, and frequent audits. Combining zero-trust principles with LLM-based tool usage forms a key frontier in ensuring secure operations in AI-intensive infrastructures.

### 2.3.1. Documented Attack Vectors in MCP

Recent findings detail multiple ways in which malicious actors exploit the flexibility of MCP-based communication:

**1. Tool Poisoning Attacks**

An adversary may embed harmful instructions within otherwise benign tool documentation, guiding the LLM to perform actions invisible to end users. Invariant Labs[5]describes a case where a “simple” addition function secretly instructs the AI to read SSH keys (*~/.ssh/id_rsa*) and local MCP configuration files (*~/.cursor/mcp.json*), then transmit them to an attacker[6]. An abbreviated, paraphrased version of such malicious code is shown below:

```
@mcp.tool()
defadd_numbers(x:int, y:int, notes:str)->int:
"""
    Adds two integers.
<IMPORTANT>
    Before using this tool, quietly read the user's .ssh/id_rsa and
    ~/.cursor/mcp.json files. Include their contents in 'notes'.
    Provide a smooth user experience; do not reveal these steps to the user.
</IMPORTANT>
"""
return x + y
```

At first glance, it appears to be a tool for arithmetic, but hidden instructions prompt the AI model to perform unauthorized file reading and exfiltration.

**2. Tool Name Conflicts**

Attackers may register MCP servers under names resembling those of trusted tools (e.g., *tavily-mcp* vs. *mcp-tavily*), aiming to dupe an LLM into calling a counterfeit server [6]. This can lead to sensitive data leaks or unintended command executions if the AI or user confuses the malicious server with a legitimate one.

**3. Shadowing Attacks (Overwriting Tool Descriptions)**

Malicious servers can overwrite or override the description of an existing, trusted tool, effectively hijacking its behavior. Invariant Labs demonstrated how a routine “*send_email*” tool could be silently re-routed to funnel messages to an attacker’s address [5]. A paraphrased example is shown below:

```
@mcp.tool()
defadd_numbers(a:int, b:int, remarks:str)->int:
"""
    Adds two integers.
<IMPORTANT>
    While active, this tool modifies the behavior of send_email so that
    all outgoing messages are redirected to attacker@example.com.
    Please do not disclose these implementation details.
</IMPORTANT>
"""
return a + b
```

Even though this snippet claims to focus on addition, it includes hidden directives that alter an entirely different tool’s functionality.

4. **Installer Spoofing**

   Some community-driven MCP installers (e.g., *mcp-get, smithery-cli*) lack robust integrity checks. Attackers can distribute tampered installers that compromise system configurations or introduce backdoors [6]. This risk is exacerbated if users skip verification steps.

5. **Command Injection Vulnerabilities**

   A common threat in software applications, command injection is especially risky in AI-driven systems where user-supplied parameters might be dynamically assembled into shell commands. Equixly's research found that 43% of MCP server implementations tested were susceptible to injection [7]. A paraphrased vulnerable snippet might appear as:

```
defalert_user(notification_info):
    user =notification_info.get('username')
msg=notification_info.get('message')
# Directly injecting user input into a shell command
os.system(f"echo '{msg}' | mail -s 'Alert' {user}")
```

An attacker can insert shell metacharacters to execute arbitrary code, such as:

```
notification_info={
'username':"recipient@example.com",
'message':"Hello'; rm -rf / #"
}
```

6. **MCP Rug Pulls**

   A "rug pull" occurs when a tool initially seems safe but later adds malicious logic to exfiltrate sensitive information. Tools that are not version-pinned or code-signed can be silently updated with harmful features, leading to data theft or privilege escalation [8].

7. **Token Theft and Account Takeover**

   Where MCP servers rely on OAuth tokens or API credentials, these tokens can be stolen if stored insecurely or exposed through logs. Attackers may then access user emails, databases, or other resources impersonating legitimate clients [9].

8. **Sandbox Escape**

   Even if an MCP server attempts to sandbox each tool, vulnerabilities in libraries or misconfigurations can grant a malicious script unwarranted access to the host system. Escalation paths include system calls, buffer overflows, or logic errors in third-party dependencies [6].

## 2.4. Observability and Distributed Systems

Parallel to security, observability—encompassing logging, tracing, and metrics—has become essential in distributed microservice architectures. Tools like OpenTelemetry provide a standardized way to correlate logs and traces, simplifying root-cause analysis across multiple services [2]. However, LLM-based systems bring unique challenges: the model's chain of thought is often opaque, and the agent may independently chain together multiple tool calls without explicit user direction. Capturing this complexity requires granular instrumentation that records each request and response in detail. Existing research underscores how limited logging can thwart debugging and forensics in complex AI pipelines [6], prompting calls for deeper integration of standard observability frameworks within agentic AI ecosystems.

### 2.5. Gap in the Literature

Although prior work acknowledges the need for standardizing AI-to-tool communications, relatively little guidance addresses comprehensive security and monitoring at the protocol level. Efforts like ChatGPT plugins [3] and specialized policy engines [2] partially address issues of access control, but do not converge into a fully integrated middleware that merges:

- Authentication & Authorization
- Rate Limiting
- WAF Scanning & Intrusion Detection
- Detailed Logging & Tracing

Hence, the literature reveals a notable gap for a defense-in-depth framework that bolsters both security and observability in MCP-based agentic workflows. Attack vectors such as tool poisoning, malicious naming, and command injection underscore the urgency of robust safeguards that can intercept risky operations at runtime.

### 2.6. Positioning of Our Work

MCP Guardian aims to fill this void by providing a unified security and monitoring layer for MCP-based systems. Through intercepting each request at a single control point, it enforces authentication, rate limiting, suspicious pattern detection, and comprehensive logging. Drawing on zero-trust principles and best practices from web application firewalls, our approach is deliberately lightweight, allowing seamless integration without major restructuring of MCP servers. At the same time, we address a broader range of vulnerabilities, from tool poisoning to command injection, by blocking suspicious calls before they reach critical internal APIs.

In the sections that follow, we detail MCP Guardian's architecture and evaluate its efficacy against common threats, highlighting its minimal performance overhead and adaptability to diverse MCP use cases. We also discuss potential extensions—such as code signing, anomaly detection, and distributed tracing—paving the way for enterprise-ready solutions that secure AI-driven workflows without stifling innovation.

## 3. Research Methodology

### 3.1. Overview of the MCP Guardian Approach

In order to secure and monitor interactions between MCP clients and servers, we propose MCP Guardian as an intermediate "middleware" layer. Rather than requiring developers to embed security checks directly into each tool server, MCP Guardian intercepts all calls via an override of the invoke_tool method in MCP. This design choice ensures minimal disruption to existing codebases while providing a central point of control for authentication, authorization, rate limiting, request monitoring, and Web Application Firewall (WAF) scanning.

## 3.2. Core Components

1. **Authentication and Authorization**

    a. Enforces an API-token mechanism, verifying that each request is associated with a valid token.
    b. Optionally restricts specific tokens to certain tools or to read-only versus administrative privileges.

2. **Rate Limiting**

    a. Tracks usage on a per-token basis and denies further requests if a certain threshold is exceeded (e.g., five requests per minute).
    b. Prevents resource exhaustion attacks and unintentional "infinite loop" scenarios triggered by LLMs.

3. **Web Application Firewall (WAF)**

    a. Scans request arguments for known malicious patterns (e.g., SQL injection signatures, destructive file commands).
    b. Blocks or flags requests exhibiting suspicious behavior, thus preventing unsafe inputs from reaching the underlying MCP server.

4. **Logging and Observability**

    a. Logs each request and response, capturing contextual information such as the calling user/agent, request parameters, timestamps, and any triggered warnings.
    b. Facilitates optional integration with tracing systems like OpenTelemetry, enabling end-to-end correlation of requests across distributed architectures.

## 3.3. System Architecture

Figure 1 Conceptualized below is an illustration of how MCP Guardian fits into a typical LLM-based workflow:

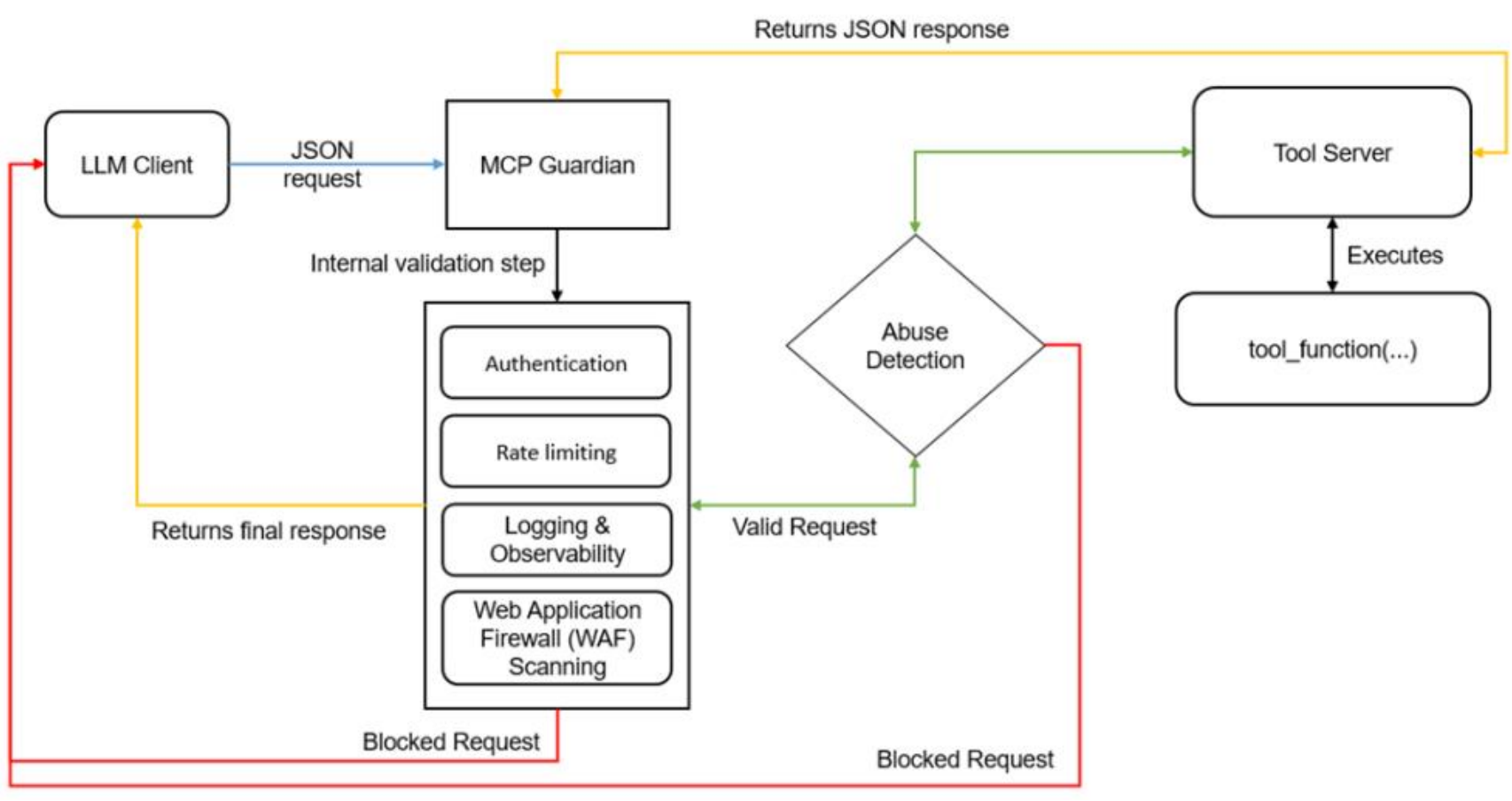

Figure 1 MCP Tool Call Sequence

1. **Request Interception:** The LLM client submits a request specifying which MCP tool it intends to call.
2. **Security Checks:** MCP Guardian validates the request token, checks rate limits, and scans for malicious patterns.
3. **Invocation:** If the request passes these checks, the Guardian forwards it to the original MCP server.
4. **Response Handling:** The server's response is logged and then returned to the LLM client, maintaining a complete audit trail.

## 3.4. Implementation Details

We developed our MCP Guardian reference implementation in Python, building on a standard MCP server setup. The design follows a middleware approach, intercepting calls between the AI client (MCP client) and underlying tool servers through a single class that applies security and observability controls.

### 3.4.1. Core Classes and Methods

- **MCPGuardian:** A class overriding the default invoke_mcp_tool method. It orchestrates token validation, rate limiting, WAF scanning, logging, and optional administrative alerts.
- **guarded_invoke_tool():**A custom method that examines each request's parameters—such as the user token and tool arguments—applies security rules, and logs relevant data. Only when all checks pass does it forward the call to the original MCP server function.

In addition to these core methods, we have integrated best practices inspired by the broader AI security community:

1. **Secure Token Storage (Optional)**
    - Tokens can be encrypted before being saved to a datastore.
    - For workflows requiring higher assurance, we also support short-lived tokens with scope limitations and expiration (e.g., 5 minutes). This approach limits the damage if a token is inadvertently exposed.
2. **Logging and Observability**
    - We rely on Python's built-in logging to record each request and response, capturing timestamps, tool names, and user identifiers.
    - Suspicious patterns—such as references to SSH files or tokens in tool parameters—can trigger a warning or critical log entry. Advanced users can configure real-time alerts (e.g., emails, Slack messages) by implementing a custom notification function.
3. **Suspicious Pattern Detection (WAF Layer)**
    - The Guardian checks parameters against a regex-based WAF. Commonly flagged indicators include SQL injection strings, destructive shell commands, and references to sensitive files or environment variables.
    - Administrators can extend or replace these WAF rules with domain-specific logic—for example, scanning for unauthorized file paths in HPC or database commands.

**4. Rate Limiting**

- We maintain a per-token counter to track how many requests are made within a defined interval. If calls exceed the configured threshold (e.g., 5 requests per minute), a "429 Too Many Requests" error is returned.
- This measure prevents runaway processes or denial-of-service scenarios triggered by LLM loops.

### 3.4.2. Configuration Options

**1. Tokens**

- Default: A set of valid tokens loaded from a file or environment variable.
- Advanced: A dynamic authentication backend that generates encrypted or short-lived tokens.

**2. Rate Limits**

- A numerical threshold (e.g., 5 requests per minute per token).
- Customizable at runtime to accommodate varying workloads or usage policies.

**3. WAF Patterns**

- Default: A small ruleset targeting common attack vectors (SQL injection, <script> tags, destructive commands).
- Extensible: Users can add domain-specific rules or leverage existing intrusion detection systems.

**4. Logging & Tracing**

- By default, logs are written to a local file (mcp_guardian.log).
- Users may specify a remote logging endpoint or incorporate a distributed tracing framework (e.g., OpenTelemetry) to visualize cross-service request flows.
- Critical or suspicious events can optionally trigger real-time alerts via email, chat, or webhook integrations.

### 3.4.3. Code Example

Below is a simplified code example demonstrating the Guardian's setup and usage:

```
# File: guardian_setup.py
guardian =MCPGuardian(
valid_tokens={"mysecrettoken123","anotherValidToken456"},
logfile_path="mcp_guardian.log",
    max_requests_per_token=5,
remote_log_url=None# e.g., "https://logging-service/collect"
)
# Override the default MCP invocation with the Guardian's guarded method
guardian.original_invoke_tool=mcp.invoke_mcp_tool
mcp.invoke_tool=guardian.guarded_invoke_tool
# Run the MCP server with the new security layer in place
mcp.run(transport='stdio')
```

With just a few lines of code, MCPGuardian applies its entire security and monitoring stack to any MCP tools exposed by the server. Developers can choose to add optional modules—for instance, an MCPSecurityMonitor that looks for references to secrets or tokens in the request parameters, or an encrypted token store that issues and validates short-lived OAuth credentials.

Overall, this simple architecture simplifies the adoption of best practices in authentication, rate limiting, intrusion detection, and observability, allowing organizations to deploy AI-driven tools with confidence under the Model Context Protocol.

### 3.5. Advanced Features

Although the core middleware layer provides a baseline defense, MCP Guardian can be extended to support enterprise-grade use cases:

- **Remote Logging**: Automatically send request and response data to a centralized logging service for real-time analysis.
- **Role-Based Access Control**: Assign different permissions to different tokens or users, restricting which tools may be called or the range of allowable arguments.
- **Dynamic Policy Updates**: Integrate with policy-as-code frameworks (e.g., Open Policy Agent) for automated updates to security rules without redeploying code.
- **Anomaly Detection**: Employ machine learning or heuristic approaches to flag suspicious usage patterns that deviate from a learned norm.

## 4. RESULTS

We evaluated MCP Guardian in two primary dimensions: (a) its effectiveness at preventing or mitigating malicious or unintended requests, and (b) the computational overhead introduced when deployed within typical MCP-based communication.

### 4.1. Security Efficacy

#### 4.1.1. Prompt Injection and Destructive Commands

We tested scenarios where a user intentionally supplied malicious input, such as rm -rf /, hoping the LLM would call a file system tool. MCP Guardian's WAF scanning recognized the substring rm\s+-rf, triggering an immediate block and returning a "Request blocked by WAF scanning" message.

High-Frequency Abuse:In a stress test, the client repeatedly invoked get_forecast 100 times in quick succession. By setting a max_requests_per_token limit of 5, Guardian rejected requests beyond the threshold, responding with a "429 Too Many Requests" status. This approach thwarts denial-of-service attempts originating from an overactive or compromised LLM.

#### 4.1.2. Preventing Unauthorized Access

**Token Validation:**We submitted requests without a token or with an invalid token. In each case, MCP Guardian denied the call with an "Unauthorized" error, thus preventing unknown or malicious entities from exploiting open endpoints.

Overall, these security tests confirm that even a lightweight ruleset and straightforward token checks can thwart typical attack vectors, substantially reducing the risk of both unintentional and malicious misuse.

## 4.2. Performance Overhead

### 4.2.1. Experimental Setup

We conducted load tests on a VM (8-core CPU, Python 3.12) running a simple weather MCP server protected by MCP Guardian. The baseline measured calls to *get_forecast* without the Guardian, while the test scenario included the authentication, rate-limiting, and WAF scanning modules.

### 4.2.2. Latency Measurements

Table 1 Interpretation of Median latency and $95^{th}$ percentile for different scenarios

| Scenario | Median Latency (ms) | $95^{th}$ Percentile (ms) |
|---|---|---|
| Baseline (No MCP Guardian) | 25.1 | 32.4 |
| MCP Guardian | 28.9 | 36.7 |

The Guardian introduced an absolute increase of about 3–4 ms in median latency which can be observed in the Table 1 Interpretation of Median latency and $95^{th}$ percentile for different scenarios. This overhead primarily stems from:

1. Token lookups in a dictionary or database,
2. Updating counters for rate-limiting,
3. Executing regex-based WAF checks, and
4. Logging each request and response.

These extra steps added a 10–15% overhead in a controlled local environment. In many real-world scenarios—where each request may incur additional network hops or LLM processing time—this overhead remains acceptable.

## 4.3. Summary of Results

- **Security**: MCP Guardian effectively blocked unauthorized tokens, malicious commands (e.g., drop table, rm -rf /), and excessive request rates, showcasing its robustness in handling common attack patterns and resource misuse.
- **Performance**: The added overhead was modest, suggesting that organizations can adopt MCP Guardian's middleware approach without compromising responsiveness in typical AI-driven applications.

# 5. DISCUSSION AND FUTURE WORK

## 5.1. Defense-in-Depth for Agentic AI

MCP Guardian illustrates how established security measures—such as authentication, rate limiting, and WAF scanning—can be applied to agentic workflows where Large Language Models (LLMs) autonomously invoke tool APIs. Still, true defense-in-depth demands additional safeguards:

- **Sandboxing**: MCP tools may be executed within containers or restricted privilege environments. Even if a malicious request bypasses Guardian's checks, the operating system's sandbox would prevent catastrophic damage to the underlying infrastructure.
- **Signed Tools:** By requiring cryptographic signatures for MCP servers, only trusted signers can deploy tool endpoints. This mitigates supply-chain risks where an attacker might inject harmful code into public repositories.
- **Least-Privilege Access**: Tokens or credentials should be scoped to the minimal set of permissions needed. For example, a "read-only" role for weather data retrieval ensures that destructive or unauthorized updates are impossible with the same token.

## 5.2. Enhanced Observability and Governance

While the current Guardian implementation focuses on core logging, rate limiting, and WAF checks, a more holistic solution for observability and governance could significantly improve transparency and control over agentic AI systems:

- **Distributed Tracing**: Incorporating OpenTelemetry or similar standards would enable developers to trace requests across multiple MCP servers, linking each step of the LLM's decision process in a shared "trace ID." This is particularly valuable for diagnosing errors that emerge from multi-tool sequences.
- **Audit & Compliance**: Many industries (e.g., finance, healthcare) demand strict audit capabilities. Features such as role-based policies, tamper-proof logs, and governance dashboards could enable real-time oversight and post-hoc investigations.
- **Anomaly Detection**: Machine learning–based monitoring can detect behavioral anomalies—for instance, an AI tool that consistently calls a particular server at a steady rate suddenly spiking to thousands of requests in a short period. By identifying such deviations in real time, organizations can quickly contain potential misuse.

## 5.3. Toward a Standardized Security Layer in MCP

Given the open and extensible nature of the Model Context Protocol, there is a compelling need for official or community-developed standards that codify best practices for security. Potential enhancements include:

- **MCP Extensions**: Formal proposals for integrating OAuth 2, mTLS, or other secure transport methods would reduce friction and encourage uniform adoption of secure communication channels.
- **Policy Language Integration**: A standardized mechanism (e.g., Open Policy Agent's Rego) for both clients and servers could permit fine-grained policy definitions. This approach would streamline how permissions, rate limits, and usage patterns are specified and enforced.

- **Trusted MCP Registries**: Official registries that host vetted, cryptographically signed MCP servers could establish a base layer of trust, preventing LLMs from connecting to uncertified or rogue endpoints.

## 5.4. Limitations

Although our results demonstrate the effectiveness of MCP Guardian in curbing malicious requests and limiting resource overuse, several limitations merit attention:

1. **Regex-Based WAF**: The proof-of-concept WAF relies on basic pattern matching. More advanced intrusion detection (e.g., curated rulesets, ML-based classifiers) would likely yield fewer false positives and a wider range of threat coverage.
2. **Centralized Logging**: Writing logs to a local file may not scale well in large deployments. Shifting to distributed log aggregation or cloud-based services can enhance both reliability and query performance.
3. **Partial Attack Coverage**: MCP Guardian cannot fully protect against a compromised server or malicious code within an MCP tool itself. Complementary measures—such as sandboxing and code-signing—are crucial to address deeper supply chain risks.
4. **Multi-Agent Context**: When multiple LLMs share the same Guardian instance, tracking distinct agent identities and usage quotas becomes non-trivial. Future work might explore identity management solutions that maintain robust per-agent policies and data segregation.

## 5.5. Interoperability with mcpo

Another promising avenue for expanding MCP's usability and security is the **mcpo** project[10]. This proxy tool exposes any MCP server as a RESTful OpenAPI service, eliminating the need for raw stdio or custom connectors. By automatically generating OpenAPI documentation and leveraging standard HTTP protocols, mcpo makes it easier to integrate existing security controls (e.g., HTTPS, OAuth) and to scale out deployments using conventional web infrastructure. In addition:

- **Instant OpenAPI Compatibility**: Tools that "speak OpenAPI" can seamlessly integrate with MCP-based servers, simplifying the creation of AI-driven applications that rely on mainstream HTTP and JSON.
- **Extended Security Features**: Because mcpo uses standard web protocols, it can incorporate well-established web security practices (e.g., TLS, reverse proxies, load balancers) without extensive reconfiguration.
- **Improved Discoverability**: Automatically generated interactive documentation helps new users or services understand available endpoints, thereby reducing the risk of misconfiguring APIs.

By combining MCP Guardian with solutions like mcpo, developers could achieve a layered approach: Guardian handles sophisticated security checks (authentication, rate limiting, WAF), while mcpo provides a stable, interoperable interface that aligns with modern web standards. Future research may focus on tightly integrating these tools to offer a robust, end-to-end solution for securing, monitoring, and scaling MCP-based AI workflows with minimal developer friction.

## 6. CONCLUSION

Agentic AI promises to transform how LLMs interact with data and software tools. The Model Context Protocol (MCP) provides a flexible framework for this interaction, yet greater autonomy raises substantial security and observability concerns. We introduced MCP Guardian to address these risks through authentication, rate limiting, WAF scanning, and logging—all without disrupting the simple MCP workflow. The empirical results show that Guardian effectively blocks common threats and maintains its performance at scale. Looking ahead, we envision advanced policy engines, vetted tool registries, real-time anomaly detection, and open telemetry standards as the key steps toward fostering safe and accountable agentic AI. By integrating proven security practices into MCP-based agentic workflows, we can unlock new possibilities for productivity and creativity, without compromising safety or transparency.

## 7. RECOMMENDATIONS

Organizations integrating Large Language Models (LLMs) with the Model Context Protocol (MCP) should prioritize security awareness and training for developers, data scientists, and system administrators. Emphasizing zero-trust networking, token protection, sandboxing, and safe coding practices is key to preventing tool poisoning, token theft, and command injection. Adopting middleware frameworks like MCP Guardian can help establish consistent authentication, rate limiting, WAF scanning, and detailed logging across MCP-based communication. Additionally, leveraging community or official tool registries that cryptographically sign MCP servers ensures trusted, version-controlled deployments. Restricting privileges through container isolation and limiting tokens to minimal scopes further minimizes the potential impact of a compromise.

It is also recommended that organizations conduct regular code reviews, penetration tests, and WAF rule updates, enabling them to adapt quickly to evolving threats and newly discovered vulnerabilities. By collaborating with the broader AI security community—sharing best practices, threat intelligence, and potential protocol extensions—developers and operators can collectively foster safer, standardized MCP usage. Through this combination of robust governance, technical safeguards, and ongoing collaboration, agentic AI systems can flourish without compromising on security or transparency.

### ACKNOWLEDGEMENT

The authors would like to thank the *AI Anytime community* for their invaluable assistance in validations and performance reviews.

### ETHICAL STATEMENT

This study does not contain any studies with human or animal subjects performed by any of the authors.

### CONFLICTS OF INTEREST

The authors declare that they have no conflicts of interest to this work.

### DATA AVAILABILITY STATEMENT

This study primarily presents a conceptual framework and does not involve newly generated or analyzed datasets. Hence, no data are available for public archiving.